\documentclass{article} 

\usepackage[utf8]{inputenc}

\usepackage{iclr2018_workshop,times}
\usepackage{url}
\usepackage{graphicx}
\usepackage{multirow}
\usepackage{lipsum}
\usepackage{amsthm}
\usepackage{graphicx}

\usepackage{hyperref}
\hypersetup{
    colorlinks=true,
    urlcolor=[rgb]{0.188, 0.498, 0.886},
    linkcolor=[rgb]{0.188, 0.498, 0.886},    
    filecolor=magenta,      
    citecolor=[rgb]{0.188, 0.498, 0.886},
}

\usepackage{algorithm} 
\usepackage{program} 
\usepackage{algorithmic}
\usepackage{listings}
\usepackage{geometry}

\usepackage{caption,setspace}
\DeclareCaptionFormat{listing}{\rule{\dimexpr0.9\columnwidth+17pt\relax}{0.4pt}\par\vskip1pt#1#2#3}
\makeatletter
\newenvironment{code}[1][htb]{%
    \renewcommand{\ALG@name}{Exemplary robot scenario}
   \begin{algorithm}[#1]%
  }{\end{algorithm}}
\makeatother

\DeclareMathVersion{sans}
\SetSymbolFont{operators}{sans}{OT1}{cmbr}{m}{n}
\SetSymbolFont{letters}  {sans}{OML}{cmbrm}{m}{it}
\SetSymbolFont{symbols}  {sans}{OMS}{cmbrs}{m}{n}

\lstnewenvironment{sflisting}[1][]
  {\lstset{#1}\mathversion{sans}}{}

\usepackage[normalem]{ulem}

\title{Introducing the \textit{Robot Security Framework} (RSF), a standardized methodology to perform security assessments in robotics}

\author{
  Víctor Mayoral Vilches,
  \textbf{Laura Alzola Kirschgens},
  \textbf{Asier Bilbao Calvo}\\
  \textbf{Alejandro Hernández Cordero},
  \textbf{Rodrigo Izquierdo Pisón},
  \textbf{David Mayoral Vilches}\\
  \textbf{Aday Muñiz Rosas},
  \textbf{Gorka Olalde Mendia},
  \textbf{Lander Usategi San Juan} \\
  \textbf{Irati Zamalloa Ugarte} and
  \textbf{Endika Gil-Uriarte}
   \\
  Alias Robotics S.L. \\
  Vitoria-Gasteiz, Álava, Spain \\
  \texttt{victor@aliasrobotics.com} \\
  \And
  \textbf{Erik Tews} and
  \textbf{Andreas Peter} \\
  Services, Cybersecurity, and Safety Group \\
  University of Twente\\
  Netherlands \\
}

\usepackage{pdfpages}

\begin{document}
\maketitle

\vspace{-1em}
\begin{abstract}
Robots have gained relevance in society, increasingly performing critical tasks. Nonetheless, robot security is being underestimated. Robotics security is a complex landscape, which often requires a cross-disciplinary perspective to which classical security lags behind. To address this issue, we present the Robot Security Framework (RSF), a methodology to perform systematic security assessments in robots. We propose, adapt and develop specific terminology and provide guidelines to enable a holistic security assessment following four main layers (Physical, Network, Firmware and Application). We argue that modern robotics should regard as equally relevant internal and external communication security. Finally, we advocate against "security by obscurity". We conclude that the field of security in robotics deserves further research efforts.

\end{abstract}



\section{Introduction}
\label{sec:intro}

Robots are going mainstream. From assistance and entertainment robots used in homes, to those working in assembly lines in industry and all the way to those deployed in medical and professional facilities. For many, robotics is called to be the next technological revolution. Yet, similar to what happened at the dawn of the computer or the mobile phone industries, there is evidence suggesting that security in robotics is being underestimated. Even though the first dead human from a robot happened back in 1979  \cite{threat0}, the consequences of using these cyber-physical systems in industrial manufacturing  \cite{threat1, threat2}, professional  \cite{threat4, threat5} or commercial  \cite{threat3} environments are still to trigger further research actions in the robotics security field.


Over the last 10 years, the domains of security and cybersecurity have been substantially democratized, attracting individuals to many sub-areas within security assessment. According to recent technical reports summarizing hacker activity per sector   \cite{hackerreport2017, hackerreport2018}, most security researchers are currently working assessing vulnerabilities in websites (70.8\%), mobile phones (smartphones, 5.6\%) and Internet of the Things (IoT) devices (2.6\%), amongst others. Notwithstanding the relevance of robot vulnerabilities for most sectors of application, no formal study has yet published relevant data about robotics nor seems to be an active area of research. We believe that the main reasons for this gap are twofold. In a first aspect, security for robots is a complex subject from a technological perspective. It requires an interdisciplinary mix of profiles, including security researchers, roboticists, software engineers and hardware engineers. In a second aspect and to the best of our knowledge, there are few guidelines, tools and formal documentation to assess robot security  \cite{sedgewick2014framework, nistframework}. Overall, robot security is an emerging challenge that needs to be addressed immediately.

In an attempt to provide a solution for the second problem, this paper presents the Robot Security Framework (RSF), a systematic methodology for performing security assessments in robotics. We argue that security, privacy and safety in robotic systems should clearly be recognized as a major issue in the field. Our framework proposes a standardized methodology to identify, classify and report vulnerabilities for robots within a formal operational protocol. Throughout the description of the RSF, we present exemplary scenarios where robots are subject to the security issues hereby exposed.

The sections below are organized as follows: Section \ref{sec:previous} describes previous work in the field of security for robots. Section \ref{sec:rsf} elaborates upon the proposed framework. Finally, Section \ref{sec:conclusions} draws major conclusions out of our work.

\section{Previous work}
\label{sec:previous}

Robot security is becoming a concern that extends rapidly. However, to date, and as already briefed in the Introduction section, there are few honest and laudable efforts that elaborate into methodologies for analyzing robot's security or cybersecurity. The most relevant of those pioneering contributions is Shyvakov's work  \cite{shyvakov2017developing}. Him and collaborators  aimed to develop a preliminary security framework for robots, described from a penetration tester's perspective. The cited research piece is, to the best of our knowledge, the best piece of literature addressing robot security concerns. Nonetheless, on the basis of the content and structure of that particular work, we largely found motivation for the present work. We found extremely relevant to review, discuss, complete it, and motivate the full picture assessment from a robotics standpoint.

The author's  \cite{shyvakov2017developing} classification proposed 4 levels of security: a) physical security, b) network security, c) operating system security and d) application security. However, we find that the author lacks to some extent, the background knowledge related to the robotics field, particularly regarding the internal organization of these systems. For instance, he states that robots have "\emph{internal networks for wiring together internal components (nodes), yet, these networks miss the fact that each is a security critical element which can potentially influence the overall robot security}". Shyvakov even includes a brief category, devoted to internal networks, within his proposed framework. However, under the assumption that "\emph{normal user is usually not supposed to connect to the internal network}", he advises that of cases where "\emph{it is not possible to implement full network monitoring due to hardware limitations} but provides no further details on the rationale. By claiming that \emph {At least there should be a capability to detect new unauthorized devices on the network}" he suggests the idea that dedicated robot network security is needed. Moreover, the author discusses that "\emph{thresholds on IDS of the internal network should be lower than on the external network}" but provides no additional foundation for such a claim. We argue that such approach would lead to an incomplete security framework by obscurity. We also believe that modern robotics should converge towards enforcing identical security levels on both inner and outer communication interfaces. Therefore, we advocate for an holistic approach to robot security on the communications level into which we will elaborate.


In an attempt of providing real use-case scenarios, the author  \cite{shyvakov2017developing} recommends a preliminary implementation of the framework and provides exemplification for real robots, yet this particular part of his work remains hidden or \emph{sanitized}. Even if the reasons behind this to be kept confidential may include the interest of robot manufacturers or stakeholders, it does little favour for actual enforcement of any security framework. Therefore, we find it necessary to provide illustrative real \emph{public} cases whereto any framework may be applied.

Other contributions to robot security, have primarily focused upon providing only partial contributions, e.g. hardening particular aspects of robots, such as middleware  \cite{Dieber:2017:SRO:3165321.3165569}, and elaborated on further efforts towards the application aspect  \cite{ApplicationSecROS} or the lower communication aspects  \cite{SecurecomROS}.


Recently, some pieces of research such as  \cite{park2017security} have brought focus onto the necessity of a framework for the evaluation of IoT device security. Such existing frameworks were targeted by  \cite{shyvakov2017developing} and duly criticized as not suitable due to incompleteness. We share the view that IoT frameworks are not applicable nor valid to provide guidance into the assessment of security to the robotics landscape. It is a common misconception that robots are a particular subset of IoT devices. Due to the fact that robots are often orders of magnitude more complex than common IoT devices, robots are to be considered, if any, a sophistication of a "network of computers", consisting of a distributed logic working in an array of sensors, actuators, power mechanisms, user interfaces and other modules that have particular connectivity and modularity requirements. Other recent researches such as  \cite{2018arXiv180504101G} claim to perform structured security assessment of a particular IoT robot. Yet, all these aforementioned pieces of research remain, in our opinion, very partial and not stablishing the. Therefore, we find it necessary to systematize assessment by further elaborating on a common and universal reference procedure for robotic systems. 

Inspired by the current state of the art, \emph{inter alia}  \cite{ApplicationSecROS, SecurecomROS, Dieber:2017:SRO:3165321.3165569, shyvakov2017developing}, we propose the subsequent Robot Security Framework (RSF). We also extend the initial ideas presented in prior art and add our contribution from a roboticist's perspective. Our main contributions on top of previous work are:
\begin{itemize}
    \item \textbf{Reformulation of the categorization terms}. In particular, the term \emph{component} becomes \emph{aspect}. Component is a rather generic term in robotics and it typically refers to a discrete and identifiable unit that may be combined with other parts to form a larger entity  \cite{iso2017iec}. Components can be either software or hardware. Even a component that is mainly software or hardware can be referred to as a software or hardware component respectively. In order to avoid any confusions, rather than \emph{component}, the term \emph{aspect} will be used to categorize each layer within RSF.
    
    \item \textbf{Overall restructuring of the content}. The original structure of the work presented by Shyvakov  \cite{shyvakov2017developing} hinders its comprehension, specially for those more familiar with robots and their components. Therefore, we propose a \emph{layer-aspect-criteria} structure where each \emph{criteria} is analyzed in terms of its \emph{objective}, the \emph{rationale} or relevance, and the systematics of assessment or \emph{method}.

    \item \textbf{Formalized firmware \emph{layer}}. We adopt a commonly accepted definition of firmware suitable for the context of robotics: software that is embedded in robots. We apply this definition to the previous 'Firmware and Operating System layer' and generalize it simply as 'Firmware layer'. Besides the operating system, we include robot middleware as a relevant topic of assessment and group them both into \emph{Firmware}, according to the adopted definition.
    
    \item \textbf{Adoption of generic "component" and "module" terms}. As an alternative to the proposed "internal component" and "external component" terminology, we suggest the generic terms "component", as defined above, and "module". Both are commonly accepted as a \emph{component with special characteristics that facilitate system design, integration, inter-operability and re-use.} This way, we simplify the message when speaking about components\footnote{The term "internal components" can lead to misunderstandings. For some, internal components are those physically embedded within the robot exoskeleton. According to others, internal components are those that physically define each discrete and identifiable component and, thereby, should not be exposed nor taken into consideration from a security perspective. Ultimately, there is a third school of thought that classifies internal and external components based on a networking point of view, considering as "internal components" only those that are connected to the internal network (with no external interface access).}. In light of the above, we elaborate on the following notion: robots are composed by components and modules. Some of them are physically exposed and some others are not. Among the modules and components, some are part of the "internal network", thereby hidden from the outside from a network perspective, whereas others are freely accessible from the outside and thereby part of the external network.
    
    \item \textbf{Improved internal networking security model}. As pointed out above, according to our vision, modern robotics should converge towards the enforcement of identically strict security levels on both internal and external communication interfaces. Therefore, we propose changes to assess internal network security and justify them by presenting existing study cases.

    \item \textbf{Improved model for physical tampering attacks}. We include a series of \emph{aspects} and \emph{criteria} to detect physical attacks on robots.  We highlight the use of logging mechanisms, already present in most robots, in order to monitor suspicious physical changes therein.
    

    \item \textbf{Added exemplary scenarios}. Throughout the framework content we add exemplary scenarios to illustrate how our methodology helps to assess the security of existing robots.
    
    \item \textbf{We open source our work} and provide a variety of user-friendly representations to simplify its adoption. This work is available and freely accessible at \url{https://github.com/aliasrobotics/RSF} under GPLv3 license.
    
\end{itemize}

We believe that an integrated approach needs to be adopted to enable a mechanistic understanding that empowers reliable assessment of robot security. To this end, the following section describes our framework.

\section{The Robot Security Framework}
\label{sec:rsf}
We hereby propose a framework based on four \emph{layers} that are relevant to robotic systems. We subsequently divide them into \emph{aspects} considered relevant to be covered. Likewise, we provide relevant \emph{criteria} applicable for security assessment. For each of these \emph{criteria} we identify what needs to be assessed (\emph{objective}), why to address such (\emph{rationale}) and how to systematize evaluation (\emph{method}), as summarized in Figure 1.

\begin{figure}[h!]
\centering
 \includegraphics[width=\textwidth]{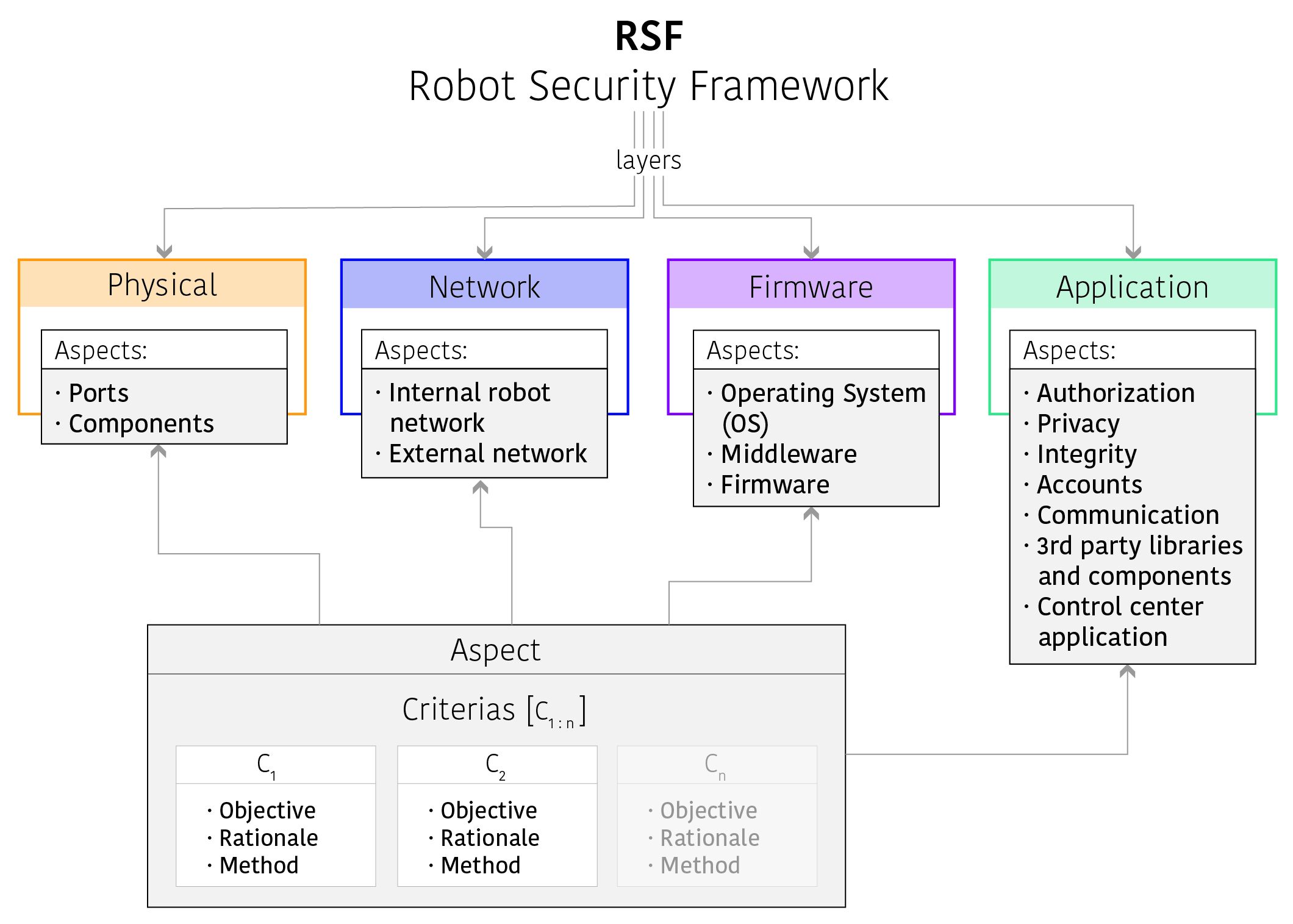}
\caption{\footnotesize {The Robot Security Framework standardized methodology: RSF is formed by 4 layers (\emph{physical, network, firmware, application}) where relevant \emph{aspects} are identified. Each aspect has different evaluation \emph{criteria} which are analyzed by three points: (1) \emph{objective} or description of the evaluation, (2) \emph{rationale} or importance of such aspect and (3) \emph{method} or systematic action plan.}}
\label{fig:robotgym}
\end{figure}

\subsection{Physical -- \emph{layer}}

\begin{figure}[h!]
\centering
 \includegraphics[width=\textwidth]{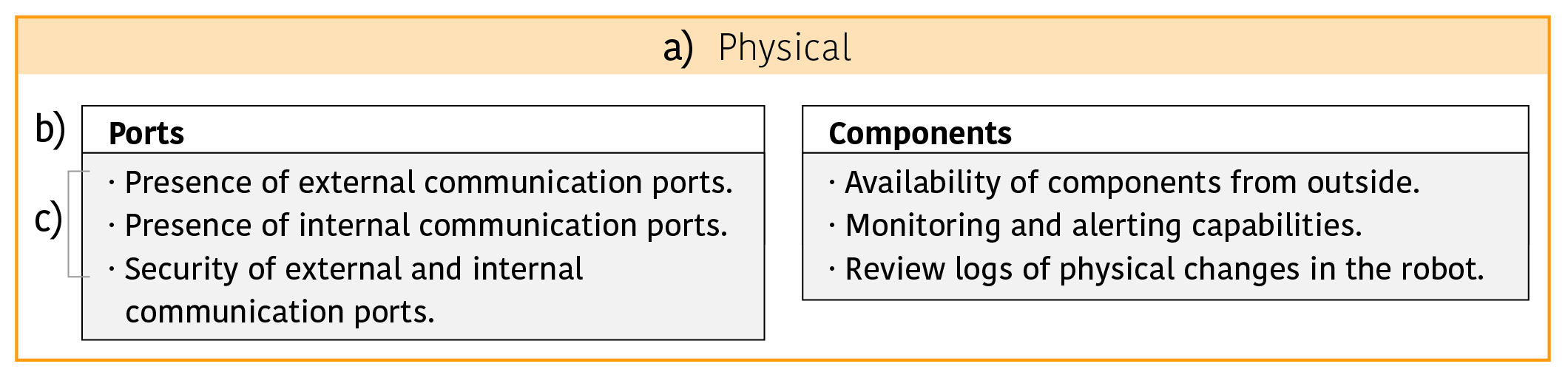}
\caption{\footnotesize The a)\textbf{ Physical} layer identifies the b) \emph{Ports} and \emph{Components} aspects, which have been analyzed with the corresponding c)criteria.}
\label{fig:robotgym}
\end{figure}
\subsubsection{Ports -- \emph{aspect}}
\begin{itemize}
    \item \textbf{Presence of external communication ports -- \emph{criteria}}
        \begin{itemize}
            \item \textbf{Objective:} Identify presence of unprotected external ports.
            \item \textbf{Rationale:} Unprotected external ports can let attackers in physical proximity to perform a variety of attacks and serve as an entry point for them.            
            \item \textbf{Method:} 
                \begin{itemize}
                    \item Inspect documentation, consult developers and inspect robot's body and components. Look for accessible ports (e.g. Ethernet, USB, CAN, etc.).
                    \item Open all doors which are not protected by locks and look for ports inside.
                \end{itemize}
        \end{itemize}
    
    \item \textbf{Presence of internal communication ports -- \emph{criteria}}\
        \begin{itemize}
            \item 
            \textbf{Objective:} Identify presence of unprotected internal ports that typically correspond with sensors, user interfaces, power or other robot-related components.
            \item
            \textbf{Rationale:} Unplugging robot components can potentially lead to the exposure of internal communication ports. Typically, these internal communication ports are not protected in robots. This may serve as an entry point and allow attackers in physical proximity to perform a variety of attacks.
            \item
            \textbf{Method:}
                \begin{itemize}
                    \item 
                     Firstly, open any door not protected by a lock. Secondly, open those protected, and look for robot components and their buses.
                    \item Investigate ventilation holes and see if they are wide enough to access internal communication ports.
                \end{itemize}
            
        \end{itemize}
\end{itemize}

\begin{code}
\caption{Physical security, exploitation of communication ports in robots}
\small  As reported by Cerrudo et al. \cite{hackingbeforeskynet, hackingbeforeskynet2}, physical attacks are possible when adversaries can access to the robot's hardware or mechanics to modify its behaviour or set up a persistent threat. Often, robots expose external ports. Such is the case of robots like Rethink Robotics' Baxter, Universal Robots' UR3  or Aldebaran Robotics' (acquired by Softbank Robotics) Pepper. In their study, Cerrudo and Apa present and demonstrate three main threats for exposed connectivity ports: USB ports, Ethernet ports and power ports. The authors present examples where each one of these threats are systematically exploited. Surprisingly, in some cases, critical damage to the robot can be caused by solely connecting a USB dongle \cite{usbhacks}.
\end{code}

\begin{itemize}        
    \item \textbf{Security of external and internal communication ports -- \emph{criteria}}\
        \begin{itemize}
            \item 
            \textbf{Objective:} Verify if attackers can sniff or modify any critical data during communication with a docking station or by connecting to the ports.
            \item
            \textbf{Rationale:} Unprotected external and internal ports can let attackers in physical proximity to perform a variety of attacks by serving as an entry point into them.
            \item
            \textbf{Method:}
                \begin{itemize}
                    \item Try to connect to the identified communication ports:
                    \begin{itemize}
                        \item
                    Determine if authentication is required (e.g. Network access control for Ethernet).
                        \item Assess whether the communication is encrypted.                \item Try communicating with them, attempt fuzzing to discover if robot’s state can be affected.
                    \end{itemize}
                    \item If a robot connects to a docking station to transfer some data, try to use sniffers to see how data exchange is being done (e.g. verify if some sensitive, configuration or control data is transferred in clear text).
                \end{itemize}
            \
        \end{itemize}
    \
\end{itemize}
\subsubsection{Components -- \emph{aspect}:}
    \begin{itemize}
        \item 
        \textbf{Availability of components from outside -- \emph{criteria}}
        \begin{itemize}
            \item 
            \textbf{Objective:}
            Identify internal hardware that is accessible from outside.
            \item
            \textbf{Rationale:} Directly accessible components can be physically damaged, stolen, tampered, removed or completely disabled causing the robot to misbehave. The most obvious example is the removal of critical sensors for the behavior of the robot.
            \item
            \textbf{Method:}
                \begin{itemize}
                    \item Inspect robots body and look for accessible components (e.g. sensors, actuators, computation units, user interfaces, power components, etc.).
                    \item Open all doors which are not protected by locks and look for accessible components inside.
                
                \end{itemize}
            \item
            \textbf{Notes:}
                 All cables should also remain inside of the robot. Some components require to be partially outside of the body frame (e.g. certain sensors such as range finders, or antennas of certain wireless communication components), in such cases only the required part should stick out, not the whole component.
        \end{itemize}
        \item
        \textbf{Monitoring and alerting capabilities -- \emph{criteria}}
            \begin{itemize}
                
                \item
                \textbf{Objective:}
            Identify whether rogue access to the internal hardware of the robot can be detected.
                \item
                \textbf{Rationale:} Having no verification whether the internals of the robot were accessed or not means that attackers can easily tamper with any components or install a hardware \emph{trojan} unnoticeably.
                \item
                \textbf{Method:}
                \begin{itemize}
                    \item Identify all parts of the frame that can be opened or removed to get access to the components or modules. 
                    \item
                Check whether there is an active (tamper switches) or passive (tamper evident screws and seals) monitoring capability present.
                    \item
                    In case of active monitoring capability, verify that the operator receives a real-time alert and the incident is being logged and acted upon by reviewing procedures.
                \end{itemize}
            \item
            \textbf{Notes:} Passive monitoring provides information upon inspection, whether internals were accessed or not. However, there is still a time window between inspections when exploited robots can be abused.
            \end{itemize}
        \item
        \textbf{Review logs of physical changes in the robot -- \emph{criteria}}
            \begin{itemize}
                \item 
                \textbf{Objective:} Verify the logs of the robot and look for tampering actions. Log examples include powering on/off events, connection/disconnection of physical components, sensor values or actuator actions. Detect potential tampering based on this information.
                \item
                \textbf{Rationale:} Most robots register logs of a variety of events going from powering on/off the robot to each individual component data. Specially, some robots detect physical changes on their components and register it. Such changes could lead to an undetected tampering of the system. Reviewing the logs could lead to discovering physical tampering of the robot.
                \item
                \textbf{Method:}
                    \begin{itemize}
                        \item Review the logs of powering on and off routines of the robot.
                        \item
                        Review the logs of physical changes in the robot.
                        \item
                        Review the logs of each individual component and look for anomalies.
                    \end{itemize}
            \end{itemize}
    \end{itemize}

\subsection{Network -- \emph{layer}}
\begin{figure}[h!]
\centering
 \includegraphics[width=\textwidth]{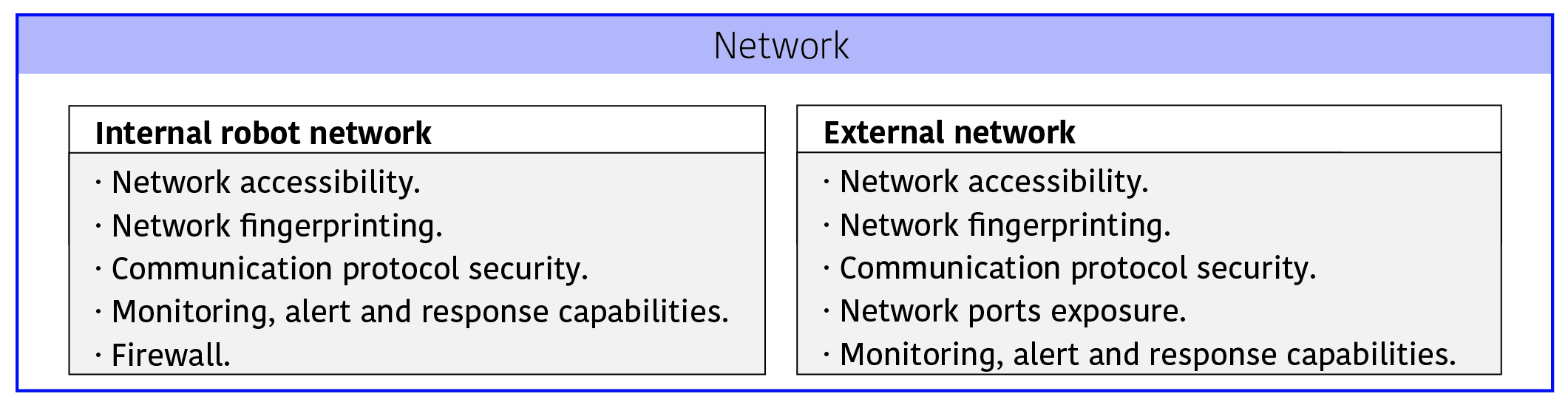}
\caption{\footnotesize The \textbf{Network} layer includes the \emph{Internal robot network} and \emph{External network} aspects, which have been analyzed with the corresponding criteria.}
\label{fig:robotgym}
\end{figure}
\subsubsection{Internal robot network -- \emph{aspect}:}
    \begin{itemize}
        \item \textbf{Network accessibility -- \emph{criteria:}}
        \begin{itemize}
            \item  \textbf{Objective:} Determine and assess network accessibility and the corresponding protection mechanisms.
            \item \textbf{Rationale:} Internal networks could be password protected. If that's the case, the corresponding mechanisms should be up-to-date and ensure that only authenticated users are able to access the network.
            \item \textbf{Method}
            \begin{itemize}
                    \item Validate authentication mechanisms and verify that no known vulnerabilities are present on such.
                    \item If internal network is password protected, attempt common password guessing.
                    \item Verify whether the robot logs both successful and unsuccessful login attempts. 
            \end{itemize}
        \end{itemize}
        
        \item \textbf{Network fingerprinting -- \emph{criteria:}}
        \begin{itemize}
            \item \textbf{Objective:} Mitigate the fingerprinting impact on the internal networks.
            \item \textbf{Rationale:}
            Network fingerprinting is useful to understand the internal network structure and its behavior, and to identify components' operating system by analyzing packets from that component. This information could be used for malicious purposes, since it provides fine-grained determination of an operating system and its characteristics.
            \item 
            \textbf{Method:}
                \begin{itemize}
                    \item Perform fingerprinting attacks on the internal networks.
                    \item Evaluate obtained information with the manufacturer's available data and assess its impact.
                    \item If necessary, propose a mitigation strategy through the use of  "scrubbers", which will "normalize" the packets, and remove the unique identifying traits that the attacker is seeking. Refer to  \cite{articlefinger} for more details about the use of "scrubbers".
                \end{itemize}
        \end{itemize}

        \item\textbf{Communication protocol security -- \emph{criteria:}}
        \begin{itemize}
            \item \textbf{Objective:} Check if used communication protocol is up-to-date, secure and has no known vulnerabilities.
            \item \textbf{Rationale:} Vulnerabilities in communication protocols can allow attackers to gain unauthorized access to the internal network of the robot and intercept or modify any transmitted data.
            \item \textbf{Method:}
            \begin{itemize}
                \item Identify all present communication capabilities by inspecting documentation, by consulting developers or by manual analysis.
                \item Analyze if used protocol versions provide encryption and mutual authentication.
                \item Verify that used protocol is hardened according to industry standards.
            \end{itemize}
        \end{itemize}

        \item 
        \textbf {Monitoring, alert and response capabilities -- \emph{criteria:}}
            \begin{itemize}
                \item 
                \textbf{Objective:}
                Identify whether internal network activity is monitored, alerts are issued and corresponding actions are taken based on known signatures or anomalies.
                \item
                \textbf{Rationale:} Proper security controls on the internal network might be challenging due to hardware limitations or performance requirements, although it is critical for robots to introspect, monitor, report and act on issues that could appear on their internal networks. Security by obscurity is unfortunately a commonly accepted approach in robotics, nevertheless, it has been demonstrated that this approach leads to critically unsecured robots. Monitoring and control capabilities should be implemented on the internal network of the robot either through the manufacturer or through additional or external solutions. The decision of applying counter-measures to the attack or only alerting should be determined by the impact of possible false positives on the operative of the robot.
                \item
                \textbf{Method:}
                    \begin{itemize}
                        \item Sweep the internal robot network and enumerate entry points (e.g. open ports, existing component information, network map of components, etc).
                        \item Try to match the fingerprints identified and to map known vulnerabilities.
                        \item Connect to the network and attempt to perform network-based attacks (e.g. ARP poisoning, denial of service on a particular component, etc.).
                        \item Verify whether the robot detects and registers incidents.
                        \item  Verify whether the robot acts upon such events and either:
                            \begin{itemize}
                                \item (The robot) responds to insult proactively.
                                \item An operator receives a real time alert and acts based on procedures.
                            \end{itemize}
                        
                    \end{itemize}
            \item
            \textbf{Notes:}
            In those cases where it is not possible to implement internal robot network monitoring, alerting and response due to limitations on the robot capabilities, manufacturers should extend their capabilities or refer to third party solutions that could offer such.
            \end{itemize}
        \item
        \textbf {Firewall -- \emph{criteria:}}
            \begin{itemize}
                \item 
                \textbf{Objective:} Identify whether internal network is separated from the external by the firewall.
                \item
                \textbf{Rationale:}
                Firewalls can help to further protect components, modules and communications from the outside and ensure that they cannot accidentally leak data to the external network.
                \item 
                \textbf{Method:}
                    \begin{itemize}
                        \item Inspect documentation, consult developers and inspect components which are responsible for external communications. Identify that such components have firewalls.
                        \item Inspect firewall settings and verify that no components or modules are allowed to communicate to the external network unless it is necessary.
                        \item If a VPN is used, verify that there are rules which allow components or modules to communicate with the outside world only via the VPN tunnel.
                    \end{itemize}
                \item \textbf{Notes:} There should be a firewall per each interface with external networks. This means that each communication component interfacing with an external channel should have a firewall behind it or use third party solutions. For example, WiFi hotspots in robots or LTE/UMTS transceivers.
                
            \end{itemize}
    \end{itemize}

\begin{code}
\caption{Internal robot network security}
\small Robots are typically built by interconnecting different components (sensors, actuators, communication devices, etc.) through internal networks. Most manufacturers, aware about the security risks, allocate resources in the protection of the interface with external networks through firewalls, IDSs or other mechanisms. To the best of our knowledge, few robotic solutions care about internal networking. Most internal networks in robotic solutions are totally unprotected, often unencrypted, and even when encrypted, authentication is unhandled, which can easily lead to Denial-of-Service (DoS) attacks.  Some pioneering pieces of research \cite{shyvakov2017developing} have reported that internal robot networks are critical elements which can influence robots security, though they fail to acknowledge the latter's relevance.

This security fault was illustrated in Cerrudo and Apa's work  \cite{hackingbeforeskynet, hackingbeforeskynet2}, where they demonstrate how Universal Robots' UR3 exposes a number of services when connected to the internal network. Neither the internal network, nor any of these services demands authentication. Thereby, any user connected to the internal network can issue commands to these services and control/rule the robot as desired.
\end{code}

\subsubsection{External network -- \emph{aspect}:}
    \begin{itemize}
        \item \textbf{Network accessibility -- \emph{criteria:}}
        \begin{itemize}
            \item  \textbf{Objective:} Determine and assess network accessibility and the corresponding protection mechanisms.
            \item \textbf{Rationale:} External networks could be password protected. If that is the case, the corresponding mechanisms should be up to date and ensure that only authenticated users are able to access the network.
            \item \textbf{Method:}
            \begin{itemize}
                    \item Validate authentication mechanisms and verify that no known vulnerabilities are present on such.
                    \item If external network is password protected, attempt common password guessing.
                    \item Verify whether the robot logs successful and unsuccessful login attempts.
            \end{itemize}
        \end{itemize}
\end{itemize}

\begin{code}
\caption{Unauthorized Modbus read and write requests }
\small  As reported by Cerrudo and Apa \cite{hackingbeforeskynet, hackingbeforeskynet2},
Universal Robots' UR3, UR5 and UR10 have a default Modbus service (port 502) that does not provide authentication of the source of a command. According to the authors, \emph{"an adversary may attempt to corrupt the robot in a state to negatively affect the process being controlled. An attacker with IP connectivity to the robot can issue Modbus write requests. This could change the state of the robot, make it interoperable, or send requests to actuators to change the state of the joints being controlled."} 
\end{code}        

\begin{itemize}
        \item \textbf{Network fingerprinting -- \emph{criteria:}}
        \begin{itemize}
            \item \textbf{Objective:} Mitigate the fingerprinting impact on the external networks.
            \item \textbf{Rationale:}
            Network fingerprinting is useful to understand the network structure and its behavior, as well as to identify devices' operating system by analyzing packets from that network. This information could be used for malicious purposes since it provides fine-grained determination of an operating system and its characteristics.
            \item 
            \textbf{Method:}
                \begin{itemize}
                    \item Perform fingerprinting attacks on the external network.
                    \item Evaluate obtained information with the manufacturer's available data and assess its impact.
                    \item If necessary, propose a mitigation strategy through the use of  "scrubbers", which will "normalize" the packets, and remove the unique identifying traits that the attacker is seeking. Refer to  \cite{articlefinger} for more details about the use of "scrubbers".
                \end{itemize}
        \end{itemize}
\end{itemize}

\begin{code}
\caption{Network fingerprinting of robots}
\small  According to Cerrudo and Apa \cite{hackingbeforeskynet, hackingbeforeskynet2},
finding certain robots in large networks is trivial and can be done using simple multicast DNS (mDNS). The authors report that these machines advertise their presence on the network through mDNS. Computers located in the same subnetwork and with support for mDNS can resolve the robot’s hostname and easily reach it. Some examples are provided below:
\begin{itemize}
    \item Softbank Robotics' Nao default hostname is "\emph{nao.local}".
    \item Softbank Robotics' Pepper default hostname is "\emph{nao.local}", also publicly available in the official documentation \cite{pepperdocs}.
    \item Rethink Robotics' Baxter and Sawyer default hostname is the serial number followed by local. E.g. "\emph{011303P0017.local}" or \emph{"$<robot name>.local$"}, also confirmed in Rethink's official documentation \cite{rethinkdocs}.
    \item Universal Robots' UR3, UR5 and UR10 default hostname is \emph{"ur.local"}.
\end{itemize}

All of them hardcoded, regardless of the device itself.
\end{code}        

\begin{itemize}
        \item\textbf{Communication protocol security -- \emph{criteria}}
        \begin{itemize}
            \item \textbf{Objective:} Check if used communication protocol is up-to-date, secure and has no known vulnerabilities.
            \item \textbf{Rationale:} Vulnerabilities in communication protocols can allow attackers to gain unauthorized access to the external network of the robot and intercept or modify any transmitted data.
            \item \textbf{Method:}
            \begin{itemize}
                \item Identify all communication capabilities being present by inspecting documentation, consulting developers or by manual analysis.
                \item Analyze if used protocol versions provide encryption and mutual authentication.
                \item Verify that used protocol is hardened according to industry standards.
            \end{itemize}
            \item \textbf{Notes}:  If providing encryption on the protocol level is not possible for some reasons, VPN or application level encryption should be used.
        \end{itemize}
\end{itemize}

\begin{itemize}
        \item \textbf{Network ports exposure -- \emph {criteria}:}
        \begin{itemize}
            \item \textbf{Objective:} Identify whether only necessary network ports are exposed to the external network.
            \item \textbf{Rationale:} More open ports mean a bigger attack surface and therefore their number should be as low as possible. Services that are exposed should have no known vulnerabilities due to the ease of their exploitation.
            \item \textbf{Method:}
            \begin{itemize}
                \item Connect to the network that is being used by the robot for communication and scan all robot ports to find the open ones. Verify in manufacturer manuals whether their presence is required.
                \item Identify, if possible, services running behind an open port and its version.
                \item Verify whether identified services are still receiving security updates and have no known vulnerabilities.
            \end{itemize}
        \end{itemize}
        
        \item \textbf{Monitoring, alert and response capabilities -- \emph{criteria}}
        \begin{itemize}
            \item \textbf{Objective:} Identify whether the external network activity is being monitored, alerts are issued based on known signatures or anomalies, and appropriate actions are taken.
            \item \textbf{Rationale:} Properly configured external network monitoring can spot network based attacks in their inception even if other security mechanisms are compromised.
            \item \textbf{Method:}
            \begin{itemize}
                \item 
                Sweep the external robot network and enumerate entry points (e.g. open ports, protocol information, network map of components, robot components etc.).
                \item Try to match the fingerprints identified and map to known vulnerabilities.
                \item Perform network based attacks (e.g. ARP poisoning, denial of service on a particular component).
                \item Verify whether the robot detects and registers incidents on the external network.
                \item Verify whether the robot acts upon such events and either:
                \begin{itemize}
                    \item (The robot) responds to insult proactively.
                    \item An operator receives a real time alert and acts based on procedures.
                \end{itemize}
            \end{itemize}
        \item \textbf{Notes:} In those cases where it is not possible to implement external robot network monitoring, alerting and response due to limitations on the robot capabilities, manufacturers should extend their capabilities or refer to third party solutions that could offer such.
        \end{itemize}
    \end{itemize}

\subsection{Firmware -- \emph{layer}}

\begin{figure}[h!]
\centering
 \includegraphics[width=\textwidth]{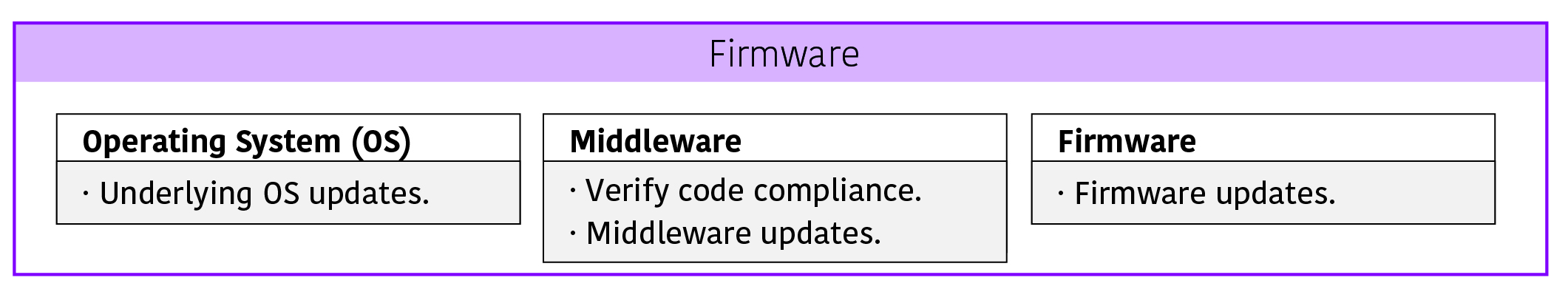}
\caption{\footnotesize The \textbf{Firmware} layer includes the \emph{Operating System, Middleware} and \emph{Firmware} aspects, which have been analyzed with the corresponding criteria.}
\label{fig:robotgym}
\end{figure}
\subsubsection{Operating System (OS) -- \emph{aspect}}
\begin{itemize}
\item \textbf{Underlying OS updates  -- \emph{criteria}}
  \begin{itemize}
    \item \textbf{Objective:} Verify that the used Operating System (OS) is still supported by the manufacturer and there is a mechanism to perform system updates.
    \item \textbf{Rationale:} Outdated operating systems can have security vulnerabilities.
    \item \textbf{Method:}
      \begin{itemize}
        \item Check if the underlying OS is still maintained and able to recieve security patches.
        \item Check whether the latest security updates are applied.
        \item Check if there is an update mechanism present and enabled.
        \item Check if updates are authenticated when transferred to the robot.
        \end{itemize}
    \end{itemize}
\end{itemize}    

\subsubsection{Middleware -- \emph{aspect}}
\begin{itemize}
\item \textbf{Verify code compliance (if accessible)  -- \emph{criteria}}
  \begin{itemize}
    \item \textbf{Objective:} In those cases where it applies (white box assessment), ensure compliance of middleware code against established compliance mechanisms.
    \item \textbf{Rationale:} As robotics and autonomy grow, especially in certain fields of robotics, users of middlewares need to be able to determine if the software is able to be used in a safety-critical environment. With suitable guidance and modification, it is expected that middleware code could be integrated as part of certain compliant system. For this purpose, code should be developed and reviewed following certain guidelines. The most common one is the Motor Industry Software Reliability Association (MISRA), widely used in many safety-critical environments and adopted by the ROS 2 middleware.
    \item \textbf{Method:}
      \begin{itemize}
        \item Determine the exact set of guidelines that are being applied.
        \item Validate whether these guidelines have been implemented.
        \end{itemize}
    \end{itemize}

\item \textbf{Middleware updates  -- \emph{criteria}}
  \begin{itemize}
    \item \textbf{Objective:} Verify that the used middleware is still maintained and supported by the manufacturer. Verify the capacity to perform system updates.
    \item \textbf{Rationale:} Outdated middlewares in robotics are subject to have security vulnerabilities. This is specially true with ROS, ROS 2 and other robot-related middlewares.
    \item \textbf{Method:}
      \begin{itemize}
        \item Check if the underlying middleware is still maintained and able to recieve security patches.
        \item Check whether the latest security updates are applied.
        \item Check if there is an update mechanism present and enabled.
        \item Check if the updates are encrypted when transferred to the robot.  
        \end{itemize}
    \end{itemize}
\end{itemize}

\subsubsection{Firmware -- \emph{aspect}}
\begin{itemize}
\item \textbf{Firmware updates  -- \emph{criteria}}
  \begin{itemize}
    \item \textbf{Objective:} Check if manufacturer firmware can be securely updated.
    \item \textbf{Rationale:} If new vulnerabilities are discovered it is important to ensure that there is a way to provide updates to all the devices that are already sold to customers. However, update mechanisms can be circumvented by an attacker to deliver malicious updates. Therefore, it is important to verify the origin of the update prior to installation.
    \item \textbf{Method:}
      \begin{itemize}
        \item Identify if there is a mechanism to deliver firmware updates.
        \item Verify whether updates are cryptographically signed.
        \item Verify that the signature is verified prior to installation.
      \end{itemize}
  \end{itemize}
\end{itemize}

\begin{code}
\caption{Insecure firmware upgrade in Nao and Pepper robots}
\small  As reported by Cerrudo and Apa \cite{hackingbeforeskynet, hackingbeforeskynet2},
Softbank Robotics' Nao and Pepper robots are subject to a firmware upgrade vulnerability due to insecure firmware upgrade mechanism. According to the authors, 
\emph{"It is possible to upgrade system components with unsigned firmware images by skipping the signature integrity check."} This vulnerability applies to Softbank's NAOqi framework, a proprietary robot programming framework used in Softbank Robotics' products. In particular, NAOqi 2.1.4.13 (NAO), NAOqi 2.4.3.28 (Pepper) and NAOqi 2.5.5.5 (Pepper) have been found vulnerable.
\end{code}


\subsection{Application -- \emph{layer}}

\begin{figure}[h!]
\centering
 \includegraphics[width=\textwidth]{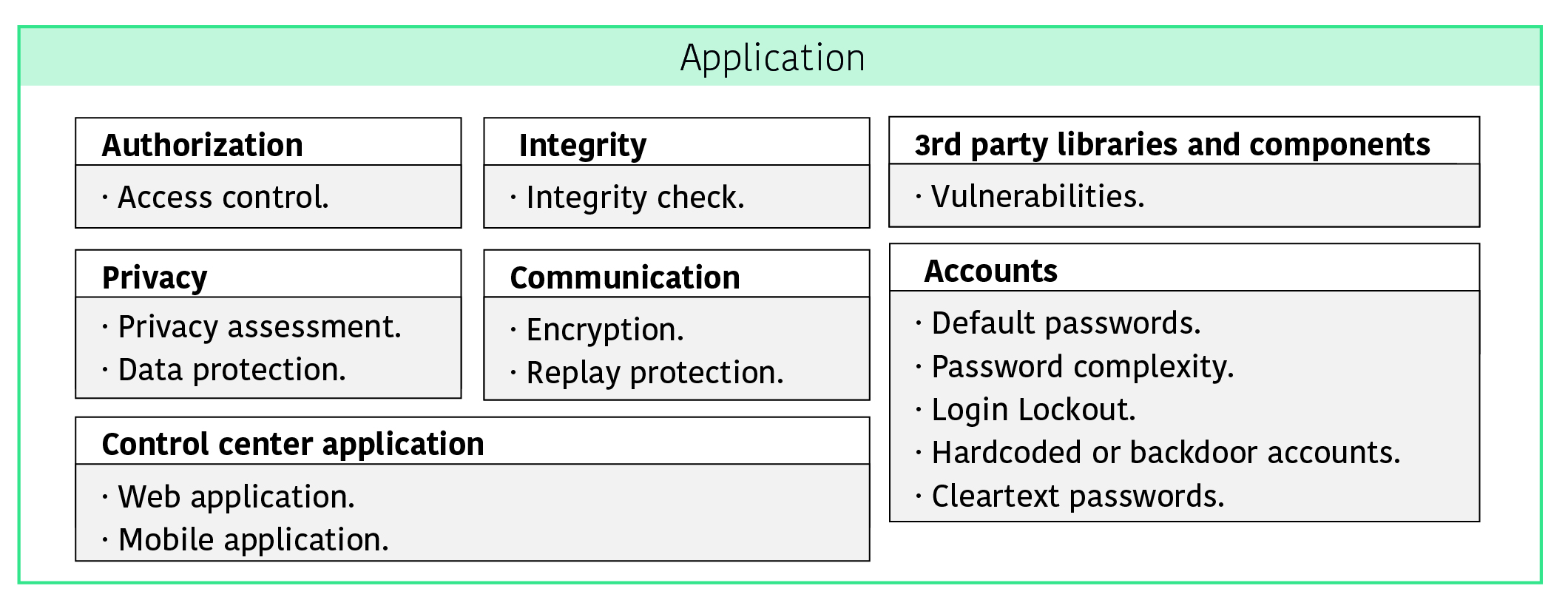}
\caption{\footnotesize The \textbf{Application} layer includes the \emph{Authorization, Privacy, Integrity, Accounts,Communication, 3rd party libraries and components} and \emph{Control center application} aspects, which have been analyzed with the corresponding \emph{criteria}.}
\label{fig:robotgym}
\end{figure}
\subsubsection{Authorization  -- \emph{aspect}}
  \begin{itemize}
    \item \textbf{Access control   -- \emph{criteria}}
    \begin{itemize}
      \item \textbf{Objective:} Verify that resources are accessible only to authorized users or services.
      \item \textbf{Rationale:} Access to the restricted functions by anonymous users or users with lower access control rights diminishes all the benefits of access control.
      \item \textbf{Method:}
        \begin{itemize}  
            \item Log in with authorized credentials and attempt to perform different actions, record the requests that are being made.
            \item Log out and attempt to send the same requests as an unauthenticated user. Verify whether it is successful.
            \item Log out and log in again as a user with lower access rights. Attempt to send the same requests again. Verify whether it is successful.
        \end{itemize}
    \end{itemize}
  \end{itemize}    

\subsubsection{Privacy  -- \emph{aspect}}
  \begin{itemize}
    \item \textbf{Privacy assessment   -- \emph{criteria}}
    \begin{itemize}
      \item \textbf{Objective:} Identify whether the robot is compliant to the privacy policies that apply.
      \item \textbf{Rationale:} Not complying with privacy standards could result in a breach of personal data.
      \item \textbf{Method:}
      \begin{itemize}  
            \item Verify that minimum Personally Identifiable Information (PII) is collected and transmitted over the internet.
            \item Verify that if PII is collected users are made aware of it (e.g. in case of a video recording people can be warned by stickers or signs on the robot).
            \item Verify that all PII is stored and transmitted in a secure manner.
      \end{itemize}
    \end{itemize}  

    \item \textbf{Data protection   -- \emph{criteria}}
    \begin{itemize}
      \item \textbf{Objective:} Identify whether the robot manufacturer emplaces mechanisms to ensure compliance to the data protection regulations and laws that apply. Particularly, General Data Protection Regulation (GDPR) in the EU.
      \item \textbf{Rationale:} Not complying with regulations could result in penalties.
      \item \textbf{Method:}
      \begin{itemize}
            \item Assess the legitimate interests.
            \item Assess the consent.
            \item Assess the information provisions.
            \item Assess the third party data.
            \item Assess the profiling.
            \item Assess the legacy data.
      \end{itemize}
      \item \textbf{Note:} This \emph{criteria} is not relevant to the security of the robot itself. However, not complying with data protection regulations can result in financial penalties and should therefore be taken into consideration in organizations using robots.
    \end{itemize}
  \end{itemize}

\subsubsection{Integrity  -- \emph{aspect}}
  \begin{itemize}
    \item \textbf{Integrity check   -- \emph{criteria}}
    \begin{itemize}
      \item \textbf{Objective:} Identify whether the system performs an integrity check of critical components and takes action if they are not present or modified.
      \item \textbf{Rationale:} Tampering with any of the critical components can make the robot cause physical damage to people and property.
      \item \textbf{Method:}
      \begin{itemize}
            \item Consult documentation and developers to find whether integrity check for critical components is present.
            \item Try disabling or modifying critical components (e.g. safety sensors or range finding systems) of the robot and check if the operator receives a real-time alert and the incident is being logged.
            \item Check whether the robot continues to function afterwards. Its operation should be stopped as soon as any critical component is disabled or modified, (e.g. if a proximity sensor is disabled the robot should not be able to move, because it will not be able to spot obstacles and can easily do some physical damage).
      \end{itemize}
      \item \textbf{Note:} Critical components are those that can directly influence the robot's operation, functionality or safety.
    \end{itemize}                
  \end{itemize}

\subsubsection{Accounts  -- \emph{aspect}}
  \begin{itemize}
    \item \textbf{Default passwords   -- \emph{criteria}}
    \begin{itemize}
      \item \textbf{Objective:} Identify presence of default passwords.
      \item \textbf{Rationale:} Default passwords are easily accessible online and remain the most popular and effortless way to exploit internet connected devices.
      \item \textbf{Method:}
        \begin{itemize}
            \item Review documentation and consult developers to identify whether default passwords are used.
            \item Attempt to log in with commonly used passwords.
            \item If default passwords are used, verify whether their change is encouraged on the first use.
            \item If unique passwords are created on a per device basis, ensure that they are random and not in a sequential order.
        \end{itemize}
      \item \textbf{Note:} When trying commonly used passwords, beware of account lockouts and verify that there is a recovery mechanism present.
    \end{itemize}     

    \item \textbf{Password complexity   -- \emph{criteria}}
    \begin{itemize}
      \item \textbf{Objective:} Verify that password complexity is enforced.
      \item \textbf{Rationale:} Weak passwords may take little time to guess.
      \item \textbf{Method:} Attempt to change the password to a weak one and verify whether this change succeeded.
      \item \textbf{Note:} Password complexity requirements depend on the sensitivity of the application. In general, the minimum requirements that should be in place are:
        \begin{itemize}
            \item A password length of, at least, 8 characters.
            \item Enforce the usage of 3 of these 4 categories:lower-case, upper-case, numbers, special characters.
        \end{itemize}
    \end{itemize}

    \item \textbf{Login Lockout   -- \emph{criteria}}
    \begin{itemize}
      \item \textbf{Objective:} Identify whether the login lockout is present.
      \item \textbf{Rationale:} Having strong and non-default passwords may not be enough. Brute force attempts should be prevented by implementing a login lockout mechanism.
      \item \textbf{Method:} Attempt to log in with incorrect credentials multiple times. Verify that the account has got a lockout.      
      \item \textbf{Note:} The lockout threshold depends on the sensitivity of the service. In general, there should be 5 login attempts or less. Prior to testing, verify that the lockout recovery mechanism is being present. Accounts can be either locked out for a specific duration of time and/or they can be recovered by physical interaction with the robot.
    \end{itemize}

    \item \textbf{Hardcoded or backdoor accounts   -- \emph{criteria}}
    \begin{itemize}
      \item \textbf{Objective:} Identify presence of hardcoded or backdoor accounts.
      \item \textbf{Rationale:} Hardcoded or backdoor credentials pose the same danger as default passwords. However, their identification is usually harder due to the need for reverse engineering or possession of the source code.
      \item \textbf{Method:}
        \begin{itemize}
            \item Consult documentation and developers to identify whether hardcoded or backdoor credentials are used.
            \item Analyze the source code for hardcoded or backdoor credentials.
        \end{itemize}
    \end{itemize}

    \item \textbf{Cleartext passwords   -- \emph{criteria}}
    \begin{itemize}
      \item \textbf{Objective:} Identify whether passwords are stored in cleartext.
      \item \textbf{Rationale:} Cleartext passwords can be leveraged by an attacker for privilege escalation or lateral movement.
      \item \textbf{Method:}
        \begin{itemize}
            \item Review the source code and documentation, consult developers and identify whether passwords are stored in a cleartext.
        \end{itemize}            
      \item \textbf{Note:} Lockout threshold depends on the sensitivity of the service. In general, there should be 5 login attempts or less.
    \end{itemize}
  \end{itemize}

\subsubsection{Communication  -- \emph{aspect}}
  \begin{itemize}
    \item \textbf{Encryption   -- \emph{criteria}}
    \begin{itemize}
      \item \textbf{Objective:} Ensure that all sensitive data is transmitted over an encrypted channel.
      \item \textbf{Rationale:} If data is transmitted in a cleartext attackers can easily gather sensitive information (e.g. credentials, audio and video streams, private data).
      \item \textbf{Method:}
        \begin{itemize}  
            \item Intercept connection between a robot and a control center application or a cloud server.
            \item Use protocol analyzer to verify that transmitted data is encrypted.
        \end{itemize}
    \end{itemize}

    \item \textbf{Replay protection   -- \emph{criteria}}
    \begin{itemize}
      \item \textbf{Objective:} Ensure that transmitted data cannot be replayed.
      \item \textbf{Rationale:} If replay protection is absent, attackers can record legitimate packets and then arbitrarily replay them to achieve desired actions.
      \item \textbf{Method:}
        \begin{itemize}  
            \item Intercept the connection between the robot and a control center application or a cloud server.
            \item Record the control or configuration packets sent to/by the robot.
            \item Attempt to replay the packets and verify whether the desired action is executed.
        \end{itemize}
    \end{itemize}  
  \end{itemize}

\subsubsection{3rd party libraries and components  -- \emph{aspect}}
  \begin{itemize}
    \item \textbf{Vulnerabilities   -- \emph{criteria}}
    \begin{itemize}
      \item \textbf{Objective:} Verify that 3rd party software components do not have known vulnerabilities.
      \item \textbf{Rationale:} It is quite common to blindly rely on 3rd party components. However they can easily introduce a vulnerability into the product where they are used.
      \item \textbf{Method:}
        \begin{itemize}  
            \item Identify which 3rd party libraries and components are used and what are their versions.
            \item Look for known vulnerabilities in the current version.
            \item Verify whether the identified component is still receiving security updates and has no unpatched vulnerabilities.
            \item Verify that the latest security updates are installed.
        \end{itemize}
    \end{itemize}
  \end{itemize}

\subsubsection{Control center application  -- \emph{aspect}}
  \begin{itemize}
    \item \textbf{Web application   -- \emph{criteria}}
    \begin{itemize}
      \item \textbf{Objective:} Perform a security assessment of the web application.
      \item \textbf{Rationale:} The robot can be indirectly compromised if the attacker exploits a web control center application.
      \item \textbf{Method:}
        \begin{itemize}  
            \item Identify web interface that is being used (hosted on the robot itself or a cloud server).
            \item Use OWASP methodology to test web application against OWASP Top 10 Web application vulnerabilities.
        \end{itemize}
    \end{itemize}

\item \textbf{Mobile
 application   -- \emph{criteria}}
    \begin{itemize}
      \item \textbf{Objective:} Perform a security assessment of the mobile application.
      \item \textbf{Rationale:} The robot can be indirectly compromised if the attacker exploits a mobile phone control center application.
      \item \textbf{Method:}
        \begin{itemize}  
            \item Identify whether the robot has a mobile app that can be used to control or interact with it.
            \item Test the application against OWASP Mobile Top 10.
        \end{itemize}
    \end{itemize}
  \end{itemize}
 
\begin{code}
\caption{Android application updates in UBTech Robotics' robots }
\small  According to Cerrudo and Apa \cite{hackingbeforeskynet, hackingbeforeskynet2},
UBTech Robotics' Alpha 1S robot is subject to a vulnerability when updates are triggered from its official Android application. The authors report that \emph{"the Alpha 1S android application does not verify any cryptographic signature when downloading and installing the APK update into the mobile device. Furthermore, due to 'App-to-Server Missing Encryption' it is possible to perform a man-in-the-middle attack in order to change the APK URL and install a customized malware on the device."}
\end{code}


\section{Conclusions and future work}
\label{sec:conclusions}
Robot privacy, integrity and security should be major concerns in a society that increasingly relies on such cyber-physical systems. Few honest efforts have been conducted to address robot security systematics. However, a deep understanding of the discipline's landscape, along with vast cross-disciplinary approaches, is crucial to provide an integrated security assessment for robotics. The research work herein presents the Robot Security Framework (RSF): a standardized methodology that enables holistic evaluation of robot security. Furthermore, RSF is provided with illustrative practical real-world examples. Following a roboticist's security approach, our contribution aims at shedding light onto the robot security scene, an area which has remained obscure insofar. Now as then, as well as for the future, we are convinced that a reliable, reproducible and systematic security assessment is a must in any modern use-case of robotics. 

Future research is envisioned with regard to constant improvement and testing of RSF. An open source template of our security framework is available at \url{http://github.com/aliasrobotics/RSF} and licensed under GPLv3.  We kindly invite security researchers, robotic researchers and analysts to review, challenge and complement our work.


\section*{Acknowledgements}

This research has been partially funded by the Basque Government, throughout the Business Development Agency of the Basque Country (SPRI) through the \emph{Ekintzaile} 2018 program. Special thanks to BIC Araba for the support provided.



\bibliography{iclr2018_workshop}
\bibliographystyle{iclr2018_workshop}


%
%
%
%
%

\end{document}